# Comment on "A mathematical theorem as a basis for the second law: Thomson's formulation applied to equilibrium"

by

Dan Solomon

Rauland-Borg Corporation
3450 W. Oakton
Skokie, IL USA

Email: dan.solomon@rauland.com

March 3, 2004

## Abstract

A. E. Allahverdyan and Th . M. Nieuwenhuizen [1] in their paper "A mathematical theorem as a basis for the second law: Thomson's formulation applied to equilibrium" present a proof of the second law of thermodynamics based on quantum mechanics. In this comment on their paper I offer a counterexample to their proof.

In their paper [1] A. E. Allahverdyan and Th. M. Nieuwenhuizen present a proof of the second law of thermodynamics as a rigorous theorem of quantum mechanics. Allahverdyan and Nieuwenhuizen consider a closed statistical system. At $t = 0$ the initial Hamiltonian is $\hat{H}_0$ and the initial equilibrium state is a Gibbs distribution so that,

$$\hat{\rho}(0) = \frac{e^{-\beta \hat{H}_0}}{Z}; \quad Z = \text{tr} e^{-\beta \hat{H}_0} \tag{1}$$

where $\beta = 1/T$ and T is the temperature. At time $t = 0$ an external interaction is turned on and then turned off at time t. W is the work put into the system during this interaction and is given by [1],

$$W = \text{tr}[\hat{H}_0 \hat{V} \hat{\rho}(0) \hat{V}^\dagger] - \text{tr}[\hat{H}_0 \hat{\rho}(0)] \tag{2}$$

where $\hat{V}$ is a unitary operator that describes the interaction. Using the mathematical identity $\text{tr}[AB] = \text{tr}[BA]$ this can be rewritten as,

$$W = \text{tr}[\hat{V}^\dagger \hat{H}_0 \hat{V} \hat{\rho}(0)] - \text{tr}[\hat{H}_0 \hat{\rho}(0)] \tag{3}$$

Allahverdyan and Nieuwenhuizen claim that W must be non-negative for all possible interactions. Since the interaction is described by $\hat{V}$ this means that W must be non-negative for all possible unitary operators $\hat{V}$. I will give an example where this is not the case.

I will examine this problem for a quantized fermion field interacting with a quantized electromagnetic field in the temporal gauge. In the temporal gauge the gauge condition is given by the relationship $A_0 = 0$ [2,3,4,5] where $A_0$ is the scalar component of the electric potential. We will work in the Schrödinger picture and use natural units so that $\hbar = c = 1$. In the Schrödinger picture the field operators are time independent and the time dependence of the quantum system is reflected in the state vectors $|\Omega(t)\rangle$. The Hamiltonian is given by,

$$\hat{H}_0 = \hat{H}_{0,D} + \hat{H}_{0,M} - \int \hat{\mathbf{J}}(\mathbf{x}) \cdot \hat{\mathbf{A}}(\mathbf{x}) d\mathbf{x} \tag{4}$$

where,

$$\hat{H}_{0,D} = \frac{1}{2}\int \left[\hat{\psi}^\dagger(\mathbf{x}), H_{0,D}\hat{\psi}(\mathbf{x})\right] d\mathbf{x}; \quad H_{0,D} = -i\boldsymbol{\alpha} \cdot \nabla + \beta m \tag{5}$$

$$\hat{H}_{0,M} = \frac{1}{2}\int \left(\hat{\mathbf{E}}^2 + \hat{\mathbf{B}}^2\right) d\mathbf{x}; \quad \hat{\mathbf{B}}(\mathbf{x}) = \nabla \times \hat{\mathbf{A}}(\mathbf{x}) \tag{6}$$

$$\hat{\mathbf{J}}(\mathbf{x}) = \frac{q}{2}\left[\hat{\psi}^\dagger(\mathbf{x}), \boldsymbol{\alpha}\hat{\psi}(\mathbf{x})\right] \tag{7}$$

In the above expressions m is the fermion mass, $\boldsymbol{\alpha}$ and $\beta$ are the usual 4x4 matrices, q is the electric charge, $\hat{H}_{0,D}$ is the Dirac Hamiltonian, $\hat{H}_{0,M}$ is the Hamiltonian for the electromagnetic field, and $\hat{\mathbf{J}}(\mathbf{x})$ is the current operator. The Schrödinger picture time independent fermion field operators are $\hat{\psi}(\mathbf{x})$ and $\hat{\psi}^\dagger(\mathbf{x})$. The Schrödinger picture time independent field operators for the electromagnetic field are $\hat{\mathbf{A}}(\mathbf{x})$ and $\hat{\mathbf{E}}(\mathbf{x})$. The electromagnetic field operators are real so that $\hat{\mathbf{A}}^\dagger(\mathbf{x}) = \hat{\mathbf{A}}(\mathbf{x})$ and $\hat{\mathbf{E}}^\dagger(\mathbf{x}) = \hat{\mathbf{E}}(\mathbf{x})$.

The field operators obey the following relationships [3,4],

$$\left[\hat{A}^i(\mathbf{x}), \hat{E}^j(\mathbf{y})\right] = -i\delta_{ij}\delta^3(\mathbf{x}-\mathbf{y}); \quad \left[\hat{A}^i(\mathbf{x}), \hat{A}^j(\mathbf{y})\right] = \left[\hat{E}^i(\mathbf{x}), \hat{E}^j(\mathbf{y})\right] = 0 \tag{8}$$

and

$$\left\{\hat{\psi}_a^\dagger(\mathbf{x}), \hat{\psi}_b(\mathbf{y})\right\} = \delta_{ab}\delta(\mathbf{x}-\mathbf{y}); \quad \left\{\hat{\psi}_a^\dagger(\mathbf{x}), \hat{\psi}_b^\dagger(\mathbf{y})\right\} = \left\{\hat{\psi}_a(\mathbf{x}), \hat{\psi}_b(\mathbf{y})\right\} = 0 \tag{9}$$

where "a" and "b" are spinor indices. In addition, all commutators between the electromagnetic field operators and fermion field operators are zero, i.e.,

$$[\hat{\mathbf{A}}(\mathbf{x}),\hat{\psi}(\mathbf{y})]=[\hat{\mathbf{E}}(\mathbf{x}),\hat{\psi}(\mathbf{y})]=[\hat{\mathbf{A}}(\mathbf{x}),\hat{\psi}^\dagger(\mathbf{y})]=[\hat{\mathbf{E}}(\mathbf{x}),\hat{\psi}^\dagger(\mathbf{y})]=0 \quad (10)$$

Define the quantity,

$$\hat{G}(\mathbf{x})=\nabla\cdot\hat{\mathbf{E}}(\mathbf{x})-\hat{\rho}(\mathbf{x}) \quad (11)$$

where the current operator $\hat{\rho}(\mathbf{x})$ is defined by,

$$\hat{\rho}(\mathbf{x})=\frac{q}{2}\left[\hat{\psi}^\dagger(\mathbf{x}),\hat{\psi}(\mathbf{x})\right] \quad (12)$$

All valid state vectors $|\Omega\rangle$ must satisfy the Gauss constraint [2],

$$\hat{G}(\mathbf{x})|\Omega\rangle=0 \quad (13)$$

The unitary operator $\hat{V}$ converts the state vector $|\Omega\rangle$ into the state vector $\hat{V}|\Omega\rangle$. The Gauss constraint must be still be satisfied so that in the temporal gauge $\hat{V}$ is restricted to those operators which satisfy,

$$\hat{G}(\mathbf{x})\hat{V}|\Omega\rangle=0 \text{ for all } |\Omega\rangle \text{ for which } \hat{G}(\mathbf{x})|\Omega\rangle=0 \quad (14)$$

I will show that for the Hamiltonian operator defined in (4) it is possible to find a unitary operator $\hat{V}$ which satisfies the Gauss constraint for which W of equation (3) is negative. This is contrary to the assertion of reference [1].

Now let,

$$\hat{V}=\hat{U}\hat{R} \quad (15)$$

where $\hat{U}$ and $\hat{R}$ are unitary operators. Let $\hat{U}$ satisfy the Gauss constraint (14) and define $\hat{R}$ by,

$$\hat{R} = e^{i\hat{U}^\dagger \hat{C} \hat{U}} \tag{16}$$

where

$$\hat{C} = \int \hat{\mathbf{E}} \cdot \nabla \chi \, d\mathbf{x} \tag{17}$$

where $\chi(\vec{x})$ is an arbitrary real valued function. Note that $\hat{C} = \hat{C}^\dagger$ so that $\hat{R}^\dagger = e^{-i\hat{U}^\dagger \hat{C} \hat{U}}$ therefore $\hat{R}^\dagger \hat{R} = 1$ so that $\hat{R}$ is unitary.

Using the commutator relations we obtain,

$$\left[\hat{H}_{0,D}, \hat{C}\right] = \left[\hat{\mathbf{J}}(\mathbf{x}), \hat{C}\right] = \left[\rho(\mathbf{x}), \hat{C}\right] = 0 \tag{18}$$

and,

$$\left[\hat{\mathbf{A}}(\mathbf{x}), \hat{C}\right] = -i\nabla \chi(\mathbf{x}) \tag{19}$$

Use this result to obtain,

$$\left[\hat{\mathbf{B}}(\mathbf{x}), \hat{C}\right] = \nabla \times \left[\hat{\mathbf{A}}(\mathbf{x}), \hat{C}\right] = -i\nabla \times \nabla \chi(\mathbf{x}) = 0 \tag{20}$$

Therefore,

$$\left[\hat{H}_{0,M}, \hat{C}\right] = 0 \tag{21}$$

From the above we obtain,

$$\left[\hat{G}(\mathbf{x}), \hat{C}\right] = 0 \tag{22}$$

Next we want to determine if $\hat{V} = \hat{U}\hat{R}$ satisfies the Gauss constraint. To do this we use the obvious relationship,

$$\left[\hat{U}^\dagger \hat{O}_1 \hat{U}, \hat{U}^\dagger \hat{O}_2 \hat{U}\right] = \hat{U}^\dagger \left[\hat{O}_1, \hat{O}_2\right] \hat{U} \tag{23}$$

where $\hat{O}_1$ and $\hat{O}_2$ are operators. Use this in (22) to obtain,

$$\left[\hat{U}^\dagger\hat{G}(\mathbf{x})\hat{U},\hat{U}^\dagger\hat{C}\hat{U}\right]=\hat{U}^\dagger\left[\hat{G}(\mathbf{x}),\hat{C}\right]\hat{U}=0 \tag{24}$$

Therefore $\hat{V}=\hat{U}\hat{R}$ satisfied Gauss constraint since $\hat{U}$ has been assumed to satisfy this constraint and $\hat{U}^\dagger\hat{C}\hat{U}$, and thereby $\hat{R}$, commutes with $\hat{U}^\dagger\hat{G}(\mathbf{x})\hat{U}$.

Next we want to evaluate $\hat{V}^\dagger\hat{H}_0\hat{V}$. Use (4) to obtain,

$$\hat{V}^\dagger\hat{H}_0\hat{V}=\hat{V}^\dagger\left(\hat{H}_{0,D}+\hat{H}_{0,M}-\int\hat{\mathbf{J}}(\mathbf{x})\cdot\hat{\mathbf{A}}(\mathbf{x})d\mathbf{x}\right)\hat{V} \tag{25}$$

We will require the Baker-Campell-Hausdorff relationships [6] which states that,

$$e^{+\hat{O}_1}\hat{O}_2 e^{-\hat{O}_1}=\hat{O}_2+\left[\hat{O}_1,\hat{O}_2\right]+\frac{1}{2}\left[\hat{O}_1,\left[\hat{O}_1,\hat{O}_2\right]\right]+\ldots \tag{26}$$

Use this along with (18), (21), and (24) to obtain,

$$\hat{V}^\dagger\left(\hat{H}_{0,D}+\hat{H}_{0,M}\right)\hat{V}=\hat{U}^\dagger\left(\hat{H}_{0,D}+\hat{H}_{0,M}\right)\hat{U} \tag{27}$$

Next we obtain,

$$\hat{V}^\dagger\left(\hat{\mathbf{J}}(\mathbf{x})\cdot\hat{\mathbf{A}}(\mathbf{x})\right)\hat{V}=\left(\hat{V}^\dagger\hat{\mathbf{J}}(\mathbf{x})\hat{V}\right)\cdot\left(\hat{V}^\dagger\hat{\mathbf{A}}(\mathbf{x})\hat{V}\right) \tag{28}$$

Use (10), (19), (24), and (26) to yield,

$$\left(\hat{V}^\dagger\hat{\mathbf{J}}(\mathbf{x})\hat{V}\right)=\left(\hat{U}^\dagger\hat{\mathbf{J}}(\mathbf{x})\hat{U}\right) \tag{29}$$

and,

$$\left(\hat{V}^\dagger\hat{\mathbf{A}}(\mathbf{x})\hat{V}\right)=\hat{U}^\dagger\left(\hat{\mathbf{A}}(\mathbf{x})-\nabla\chi(\mathbf{x})\right)\hat{U} \tag{30}$$

Use these results in (28) to obtain

$$\begin{aligned}\hat{V}^\dagger\left(\hat{\mathbf{J}}(\mathbf{x})\cdot\hat{\mathbf{A}}(\mathbf{x})\right)\hat{V}&=\left(\hat{U}^\dagger\hat{\mathbf{J}}(\mathbf{x})\hat{U}\right)\cdot\hat{U}^\dagger\left(\hat{\mathbf{A}}(\mathbf{x})-\nabla\chi(\mathbf{x})\right)\hat{U}\\&=\hat{U}^\dagger\left(\hat{\mathbf{J}}(\mathbf{x})\cdot\left(\hat{\mathbf{A}}(\mathbf{x})-\nabla\chi(\mathbf{x})\right)\right)\hat{U}\end{aligned} \tag{31}$$

Use this and (27) in (25) to obtain,

$$V^\dagger \hat{H}_0 V = \hat{U}^\dagger \left( \hat{H}_0 - \int \hat{\mathbf{J}}(\mathbf{x}) \cdot \nabla \chi(\mathbf{x}) d\mathbf{x} \right) \hat{U} \tag{32}$$

Use this in (3) to yield,

$$W = W_0 - \int \text{tr}\left[ \hat{U}^\dagger \hat{\mathbf{J}}(\mathbf{x}) \hat{U} \hat{\rho}(0) \right] \cdot \nabla \chi(\mathbf{x}) d\mathbf{x} \tag{33}$$

where,

$$W_0 = \text{tr}[\hat{U}^\dagger \hat{H}_0 \hat{U} \hat{\rho}(0)] - \text{tr}[\hat{H}_0 \hat{\rho}(0)] \tag{34}$$

Integrate the second term in (33) with respect to parts and assume reasonable boundary conditions to obtain,

$$W = W_0 + \int \chi(\mathbf{x}) \nabla \cdot \hat{\mathbf{J}}_{\hat{U}}(\mathbf{x}) d\mathbf{x} \tag{35}$$

where $\hat{\mathbf{J}}_{\hat{U}}(\mathbf{x})$ is defined by,

$$\hat{\mathbf{J}}_{\hat{U}}(\mathbf{x}) = \text{tr}[\hat{U}^\dagger \hat{\mathbf{J}}(\mathbf{x}) \hat{U} \hat{\rho}(0)] \tag{36}$$

Note that $W_0$ and the quantity $\hat{\mathbf{J}}_{\hat{U}}(\mathbf{x})$ are independent of $\chi(\mathbf{x})$. Therefore $\chi(\mathbf{x})$ can be varied in an arbitrary manner without affecting $W_0$ or $\hat{\mathbf{J}}_{\hat{U}}(\mathbf{x})$. Assume that $\hat{U}$ is chosen so that $\nabla \cdot \hat{\mathbf{J}}_{\hat{U}}(\mathbf{x})$ is non-zero. In this case we can always find a $\chi(\mathbf{x})$ which makes W a negative number. For example let $\chi(\mathbf{x}) = -\lambda \nabla \cdot \hat{\mathbf{J}}_{\hat{U}}(\mathbf{x})$ where $\lambda$ is a constant. In this case (35) becomes,

$$W = W_0 - \lambda \int \left( \nabla \cdot \hat{\mathbf{J}}_{\hat{U}}(\mathbf{x}) \right)^2 d\mathbf{x} \tag{37}$$

Therefore, since $\nabla \cdot \hat{\mathbf{J}}_{\hat{U}}(\mathbf{x})$ is non-zero the quantity under the integral must be positive. Therefore for sufficiently large $\lambda$ the work W will be negative.

This result depends on the assumption that we can find a unitary transformation $\hat{U}$ such that $\nabla \cdot \hat{\mathbf{J}}_{\hat{U}}(\mathbf{x})$ is non-zero. How do we know that this can be done? $\hat{\mathbf{J}}_{\hat{U}}(\mathbf{x})$ is the current averaged over the distribution after an interaction associated with the unitary operator $\hat{U}$. Therefore we must have an interaction the takes the average current from an initial equilibrium state, where the divergence of the average current is zero, to some non-equilibrium state where $\nabla \cdot \hat{\mathbf{J}}_{\hat{U}}(\mathbf{x})$ is non-zero. Since there are many interaction that can take a system into a non-equilibrium state it is reasonable to assume that a $\hat{U}$ exists.

In conclusion a counterexample has been found to the proof presented in [1] that the quantity W in equation (2) must be non-negative for any unitary operator $\hat{V}$. The counterexample is for a quantized fermion field coupled to a quantized electromagnetic field in the temporal gauge. It is shown that in this case a unitary operator $\hat{V}$ exists for which W is negative.

## **References**

1. A.E. Allahverdyan and Th. M. Nieuwenhuizen, Physica A, **305,** 542 (2002). See also arXiv:cond-mat/0110422.

2. G. Leibbrandt, Rev of Modern Phys, **59**, 1067 (1987).

3. M. Creutz, Ann. Phys., **117**, 471 (1979)

4. B. Hatfield. "Quantum Field Theory of Point Particles and Strings", Addison-Wesley, Reading, Massachusetts (1992).

5. C. Kiefer and A. Wiepf , Annals Phys., **236**, 241 (1994).

6. W. Greiner and J. Reinhardt. "Field Quantization", Springer-Verlag, Berlin (1996).